\documentclass[fleqn,usenatbib]{mnras}

\usepackage[T1]{fontenc}

\DeclareRobustCommand{\VAN}[3]{#2}
\let\VANthebibliography\thebibliography
\def\thebibliography{\DeclareRobustCommand{\VAN}[3]{##3}\VANthebibliography}

\usepackage{graphicx}	% Including figure files
\usepackage{amsmath}	% Advanced maths commands
\usepackage{amssymb}	% Extra maths symbols

\newcommand\mabs{M}
\newcommand\vol{V}
\newcommand\env{\tilde \delta}
\newcommand\envi{\tilde \delta_i}
\newcommand\envreal{\delta_b}
\newcommand\sigmapb{\sigma_{\rm PB}}
\newcommand\sigmapbi{\sigma_{\rm PB,i}}
\newcommand\sigmapbtot{\sigma_{\rm PB,tot}}
\newcommand\epcv{\varepsilon_{\rm cv}}
\newcommand\epcvi{\varepsilon_{{\rm cv}, i}}
\newcommand\epcvone{\varepsilon_{{\rm cv}, 1}}
\newcommand\epcvtwo{\varepsilon_{{\rm cv}, 2}}
\newcommand\epcveff{\varepsilon_{\rm cv, eff}}
\newcommand\snum{N_{\rm f}}
\newcommand\veff{V_{\rm eff}(\mabs)}
\newcommand\HST{HST}
\newcommand\pakidge{\texttt{galcv}}

\title[Fitting with Cosmic Variance]{A Framework for Simultaneously Measuring Field Densities and the High-z Luminosity Function}

\author[Trapp, Furlanetto, \& Yang]{
A.C. Trapp,$^{1}$\thanks{E-mail: atrapp@astro.ucla.edu}
Steven R. Furlanetto,$^{1}$
Jinghong Yang$^{1}$
\\
$^{1}$Department of Physics and Astronomy, University of California Los Angeles, CA, 90095-1562, USA \\
}

\date{Accepted XXX. Received YYY; in original form ZZZ}

\pubyear{2021}

\begin{document}
\label{firstpage}
\pagerange{\pageref{firstpage}--\pageref{lastpage}}
\maketitle

\begin{abstract}
Cosmic variance from large-scale structure will be a major source of uncertainty for galaxy surveys at $z \ga 6$, but that same structure will also provide an opportunity to identify and study dense environments in the early Universe. 
Using a robust model for galaxy clustering, we directly incorporate large-scale densities into an inference framework that simultaneously measures the high-$z$ ($z \ga 6$) UV luminosity function and the average matter density of each distinct volume in a survey.
Through this framework, we forecast the performance of several major upcoming James Webb Space Telescope (JWST) galaxy surveys. We find that they can constrain field matter densities down to the theoretical limit imposed by Poisson noise and unambiguously identify over-dense (and under-dense) regions on transverse scales of tens of comoving Mpc. 
We also predict JWST will measure the luminosity function with a precision at $z = 12$ comparable to existing Hubble Space Telescope's constraints at $z = 8$ (and even better for the faint-end slope). We also find that wide-field surveys are especially important in distinguishing luminosity function models.

\end{abstract}


%%%%%%%%%%%%%%%%%%%%%%%%%%%%%%%%%%%%%%%%%%%%%%%%%%

%%%%%%%%%%%%%%%%% BODY OF PAPER %%%%%%%%%%%%%%%%%%

%%%%%%%%%%%%%%%%%%%%%%%%%%%%%%%%%%%%%%%%%%%%%%%%%%%%%%%%%%%%%%%%
%%%%%%%%%%%%%%%%%%%%%%%%%%%%%%%%%%%%%%%%%%%%%%%%%%%%%%%%%%%%%%%%

\section{Introduction}

The first billion years of galaxy formation is about to be explored as never before. The James Webb Space Telescope (JWST), the Nancy Grace Roman Space Telescope, the Thirty Meter Telescope, the European Extremely Large Telescope, the Giant Magellan telescope, and many other next-generation observatories will open a new frontier at the beginning of structure formation in our Universe.

Perhaps the most basic observable of this era is the galaxy luminosity function, which tracks the growth of the galaxy population as a whole. The evolution of both its shape and normalization have important implications for galaxy formation scenarios, so it has been intensely studied by existing facilities \citep[see e.g.,][]{Bouwens2015, Finkelstein2015, Bowler2015, Livermore2017, Atek2018, Oesch2018, Behroozi2019, Bouwens2021, Finkelstein2021}. To date, these measurements have pinned down the abundance of relatively bright galaxies at $z \la 8$ to a reasonable precision. The results are largely consistent with simple extrapolations of galaxy physics at lower redshifts \citep[see e.g.,][]{Tacchella2013, Mason2015, Furlanetto2017, Mirocha2017}. A few bright galaxies have also been discovered at $z \ga 9$, but they are currently too rare for robust estimates of their abundance \citep{Oesch2013,Oesch2015,Bouwens2015, Ishigaki2015,Mcleod2015,Mcleod2016,Bouwens2019}.

Despite the substantial progress in understanding these galaxies over the last several years, the field is poised for a revolution with the launch of JWST and, beyond that, the Roman Telescope. The extraordinary sensitivity of these facilities will allow galaxy searches to extend both to significantly fainter sources and to higher redshifts (e.g., \citealt{Behroozi2015,Furlanetto2017,Kauffmann2020}). It will be crucial to optimize these efforts in order to constrain the luminosity function. 

\begin{figure}
    \centering
    \includegraphics[width=0.5\textwidth]{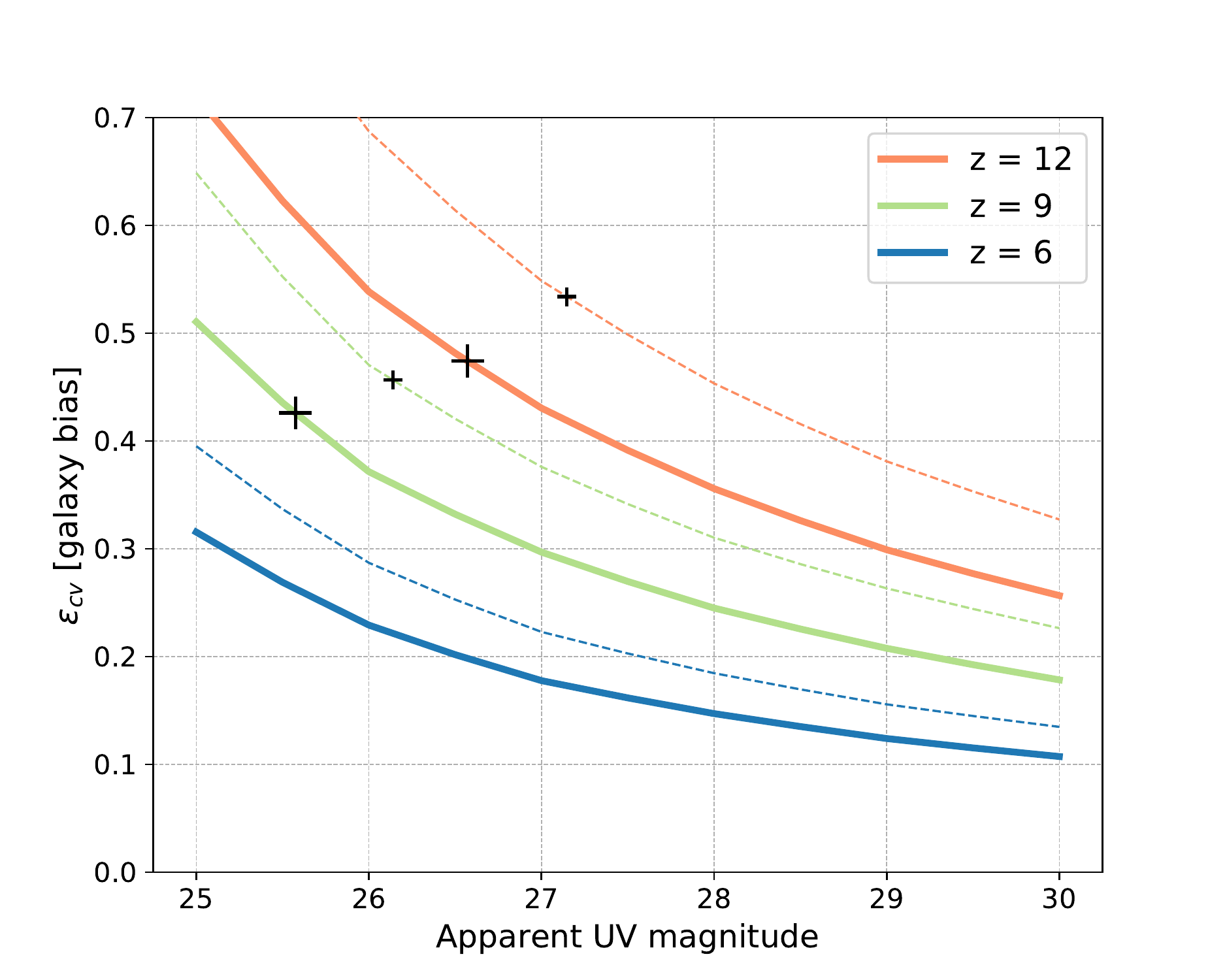}
    \caption{Galaxy bias from cosmic variance as a function of apparent magnitude for redshifts 6, 9, 12 (bottom, middle, top \textit{thick solid} lines). A galaxy bias value of e.g., $\epcv = 0.3$ means a region with a 1-$\sigma$ over-density of dark matter will have 30\% more galaxies than the Universe average for such a volume (not accounting for Poisson noise). The \textit{thick solid} lines are for a 100 arcmin$^2$ survey with $\Delta z = 1$, and the \textit{thin dashed} lines are for a 25 arcmin$^2$ survey. Galaxy bias becomes a stronger function of magnitude at higher redshift and smaller volume, meaning cosmic variance can significantly affect the shape of the luminosity function as well as the normalization. The `+' markers indicate where we would expect to find only $\sim $1 source at the indicated magnitude in such a survey, an indicator of where Poisson noise is clearly dominating.}
    \label{fig:shapechange}
\end{figure}

Going to higher redshifts comes with bigger challenges but also more opportunities. Among the obstacles is the increase in uncertainty due to cosmic variance:\footnote{In this paper, we follow common usage and apply the term ``cosmic variance'' to describe the fluctuations in dark matter density between different volumes in our Universe and the consequences of that variance for the galaxy population. More precisely, this is a particular case of sample variance, with cosmic variance sometimes reserved for the errors intrinsic to having only one Universe to observe.} large-scale dark matter densities affect the formation of dark matter haloes and thus change the normalization and shape of the luminosity function for different volumes \citep[see Figure~\ref{fig:shapechange}, and][]{Trapp2020}. 
But this cosmic variance is not simply a 
nuisance, because it reflects the large-scale structure that is itself a key driver of both galaxy formation and reionization during the Cosmic Dawn. As such, different large-scale densities (if they can be distinguished!) can serve as laboratories for discovery:
\begin{enumerate}
    \item Because ionizing sources trace the large-scale density field, the reionization process is heavily dependent on those densities \citep{Furlanetto2004}. In order to understand that era, there is great interest in identifying over-dense regions that may host large ionized bubbles or other unusual ionization environments \citep[see e.g.][]{Zitrin2015,Jung2020,Tilvi2020,Hu2021,Endsley2021} as well as large-scale underdensities that may be the last regions to be reionized \citep{Becker2018, Davies2018, Christenson2021}.
    \item The large-scale environment of galaxies appears to play a role in their evolution at later times \citep[e.g., assembly bias;][]{Gao2007}, and it may be even more important at $z \ga 6$. This is because feedback from large-scale radiation backgrounds can have enormous effects on star formation at these times. Most importantly, photoheating from reionization will suppress the formation of small galaxies inside ionized regions \citep{Thoul1996, Iliev2007, Noh2014}. Measurements of the large-scale density will allow detailed exploration of these mechanisms. 
    \item A region's density provides information on its past and future -- whether it will become a galaxy cluster, its reionization history, etc. Understanding the assembly history of unusual objects like galaxy clusters is therefore facilitated by measuring large-scale densities at early times (e.g., \citealt{Chiang2017}).
    \item Finally, while cosmologists understand the underlying dark matter density field fairly well, the transformation from those densities to galaxy observables is more uncertain. Comparing large-scale density estimates from surveys with the cosmological predictions can help identify any problems in theoretical models of galaxy formation \citep{Trapp2020}.
\end{enumerate}

In order to deal with the uncertainties of and gain insights from cosmic variance, we need a comprehensive way of modelling its effects
on the galaxy population, starting with the luminosity function and its measurement. 
In \citet{Trapp2020}, we developed a model of cosmic variance's effects on the luminosity function, including important non-linear effects and a correction for the elongated `pencil-beam' shape of many survey volumes. Recent simulations have also estimated the cosmic variance of galaxies \citep{Bhowmick2020,Ucci2021}, and those results are comparable to our analytical estimates. However, these estimates have not yet been fully integrated into luminosity function fitting methods.

A `standard method' to fit the luminosity function assumes its shape is constant across all fields, while ignoring the normalization of each one. At the end of the fitting process, a normalization is chosen such that the correct total number of sources is recovered \citep[see e.g.,][]{Finkelstein2015,Bouwens2015}.
This method ignores the dependence of the shape of the luminosity function on the density of a region at higher redshifts (see Figure~\ref{fig:shapechange}), and it discards potentially useful information about field densities.
As a result, the standard method is effective in mitigating cosmic variance, but it leads to a sub-optimal (and slightly biased) fit \citep{Trapp2020}. Here, we develop a method that integrates priors on cosmic variance into luminosity function inference, improving constraints and providing new information on the large-scale densities.

In section~\ref{sec:methods}, we use Bayesian statistics to incorporate cosmic variance into a luminosity function fitting framework from the ground up. This framework simultaneously fits luminosity function parameters and individual field densities to data from multiple fields. We also develop a method to combine multiple fields into a single `effective' field to cut down on computation time. We use our framework to forecast the performance of a selection of upcoming galaxy surveys in section~\ref{sec:applications}. 
We explore how well different combinations of the surveys can measure the luminosity function and individual field densities, finding a fundamental limit to measuring a field's density that is due to Poisson noise.
We present our conclusions in section~\ref{sec:conclusion}.

We take the following cosmological parameters: $\Omega_m = 0.308$, $\Omega_\Lambda=0.692$, $\Omega_b=0.0484$, $h=0.678$, $\sigma_8=0.815$, and $n_s=0.968$, consistent with recent Planck Collaboration XIII results \citep{PlanckCollaboration2016}. We give all distances in comoving units. All luminosities are rest-frame ultra-violet ($1500-2800$ \AA)\footnote{This wavelength range corresponds to $H$-band in the redshift range of $z\approx5$--$9$ and $K$-band for $z\approx8$--$12$.} luminosities, and all magnitudes are AB magnitudes.

%%%%%%%%%%%%%%%%%%%%%%%%%%%%%%%%%%%%%%%%%%%%%%%%%%%%%%%%%%%%%%%%
%%%%%%%%%%%%%%%%%%%%%%%%%%%%%%%%%%%%%%%%%%%%%%%%%%%%%%%%%%%%%%%%

%%%%%%%%%%%%%%%%%%%%%%%%%%%%%%%%%%%%%%%%%%%%%%%%%%%%%%%%%%%%%%%%
%%%%%%%%%%%%%%%%%%%%%%%%%%%%%%%%%%%%%%%%%%%%%%%%%%%%%%%%%%%%%%%%

\section{Methods}\label{sec:methods}

In this section, we (i) define the luminosity function, (ii) create a posterior that includes the luminosity function parameters as well as the densities of each field, (iii) develop a novel way of combining the galaxy bias functions of separate fields, and (iv) discuss the challenge of dealing with systematic offsets between datasets.

%%%%%%%%%%%%%%%%%%%%%%%%%%%%%%%%%%%%%%
\subsection{The luminosity function of galaxies}\label{sec:lf}

Let us assume that the average number of galaxies with absolute magnitudes between $(\mabs, \mabs+d\mabs)$ is described by $\Phi_{\rm avg}(\mabs,z) d\mabs$, which has parameters $\vec{\phi}(z)$ that are only functions of redshift. 
We organize the parameters of the luminosity function into a vector for notational convenience; for a typical Schechter function, these would be the normalization $\phi^*$, the characteristic magnitude $M^*$, and the faint-end power-law slope, $\alpha$:
\begin{equation}
\begin{aligned}
    &\Phi_{\rm avg}(\mabs,z) d\mabs = \\
    &(0.4~{\rm ln}10)\phi^*[10^{0.4(\mabs^*-\mabs)}]^{\alpha+1}{\rm exp}[-10^{0.4(\mabs^*-\mabs)}]d\mabs.
\end{aligned}
\end{equation}

In a given volume $\vol$, the dark matter density may differ from the average, which then makes the galaxy luminosity function dependent on that large-scale density \citep{Lacey1993}. This effect is called cosmic variance or sample variance. For large volumes, a linear approach is sufficient to describe the
conditional luminosity function \citep{Mo1996, Trapp2020}:
\begin{equation}\label{eqn:philoc}
    \Phi_{\rm loc}(\mabs,z,\env) = \Phi_{\rm avg}(\mabs,z) ( 1 + \env \epcv(\mabs,z)),
\end{equation}
where $\env$ is a normally distributed random variable signifying the \textit{normalized} relative dark matter density in the volume $\vol$. 
The actual relative density $\envreal = (\rho - \bar{\rho})/\bar{\rho}$ can be found with $\envreal = \env\sigmapb$, where $\sigmapb$ is the rms fluctuation of $\envreal$ across many volumes of shape $\vol$ at redshift $z$. Finally, $\epcv(\mabs,V,z)$ is the galaxy bias function from \cite{Trapp2020}.

The galaxy bias function $\epcv(\mabs,V,z)$ is constructed by implementing non-linear halo clustering theory into a self-consistent analytical galaxy model. The bias function also includes a correction for the elongated `pencil-beam' shape of many survey volumes and for non-linear expansion/contraction of a region. The largest uncertainty in $\epcv$ comes from non-linear halo clustering theory; $\epcv$ can vary $\sim$25$\%$ between different models. The choice of galaxy model can also affect $\epcv$, but to a lesser extent ($\epcv$ is a relatively weak function of magnitude, although it gets stronger at higher redshifts, see Figure~\ref{fig:shapechange}). See
\cite{Trapp2020} for more information on $\epcv$ and \citet{Bhowmick2020} and \citet{Ucci2021} for recent galaxy bias estimations calculated from large volume galaxy simulations.

The luminosity function that is actually measured depends on observational features like the completeness and contamination functions, which we combine into a single function $f(\mabs)$ that is unique to each survey volume. The luminosity function becomes:
\begin{equation}\label{eqn:phiobs}
    \Phi_{\rm obs}(\mabs,z,\env) = f(\mabs) \times \Phi_{\rm loc}(\mabs,z,\env).
\end{equation}
In the next section, we use a Bayesian framework to fit this observed luminosity function to data.

%%%%%%%%%%%%%%%%%%%%%%%%%%%%%%%%%%%%%%
\subsection{The posterior}\label{sec:bayes}

Given data $\vec{D}$ from a large galaxy survey composed of $\snum$ fields each with their own volume, observational effects $f(\mabs)$, and local density $\env$, we would like to learn the parameters $\vec{\phi}(z)$ of the average luminosity function. The data $\vec{D}$ is comprised of many galaxies with measured redshifts and magnitudes. Let us assume a sufficiently narrow redshift range that we may ignore any redshift dependence within the measured volume (see Appendix~\ref{sec:deltaz}).

We would like to determine the posterior $p(\vec{\phi} | \vec{D})$: the probability density of the luminosity function parameters given the data. We are also interested in $p(\vec{\delta} | \vec{D})$: the probability density of the normalized dark matter densities\footnote{$\vec{\delta}$ is a vector containing $\env$ for each survey. We omit the tilde when writing the vector for convenience.} from the $\snum$ fields given the data. A third set of parameters arise from the halo-galaxy mapping performed by a galaxy evolution model (as explored in \citealt{Trapp2020}). In this paper, we focus on understanding the large-scale density field so do not include such parameters, but we do estimate their impact on the constraints in section~\ref{sec:limitations}.

We then start then with the joint posterior $p(\vec{\phi}, \vec{\delta} | \vec{D})$. Using Bayes' theorem:
\begin{equation}
    p(\vec{\phi},\vec{\delta} | \vec{D}) \propto p(\vec{D} | \vec{\phi},\vec{\delta}) \times p(\vec{\phi}) \times p(\vec{\delta})
\end{equation}
where $p(\vec{D} | \vec{\phi},\vec{\delta})$ is the likelihood of the data given the parameters of the average luminosity function and the dark matter densities, and $p(\vec{\phi})$ and $p(\vec{\delta})$ are their respective independent priors. We assume the survey fields are separated on the sky such that their environments are independent of one another. Under that assumption, the density prior $p(\vec{\delta})$ is simply a multivariate Gaussian centered at the origin with an identity covariance matrix of size $\snum\times \snum$. We explore in section~\ref{sec:cont} how to deal with adjacent survey fields with correlated densities.

We write the total likelihood as a product of each independent field's likelihood:
\begin{equation}
    \mathcal{L} = p(\vec{D} | \vec{\phi},\vec{\delta}) = \prod_i^{\snum} p(D_i | \vec{\phi}, \envi),
\end{equation}
with $p(D_i | \vec{\phi}, \envi)$ as the likelihood that one has obtained data $D_i$ from a field with density $\envi$ and global parameters $\vec{\phi}$, which can be described by
\begin{eqnarray}\label{eq:indvlikelihood}
    & p(D_i | \vec{\phi}, \envi) & =  \\
    & & P(n_i \textrm{ galaxies are found in volume } V_i | \vec{\phi}, \envi)  \\
    & & \times p(\textrm{distribution of } \mabs | \vec{\phi}, \envi).
\end{eqnarray}
The probability that $n_i$ galaxies are found in a volume follows the Poisson distribution with expected value $n_{i,{\rm exp}}$,
\begin{equation}
    P(n_i \textrm{ galaxies are found in volume } V_i | \vec{\phi}, \envi) = \frac{n_{i,\rm exp}^{n_i}}{n_i!} e^{-n_{i,\rm exp}}
\end{equation}
where
\begin{equation}
    n_{i,\rm exp} = V_i \times \int_{-\infty}^{M_{\rm lim}} f_i(M')\Phi _{\rm avg}(M',\vec{\phi})(1+\envi\epcvi(M',V_i))\textrm{d}M'.
\end{equation}
The probability to find a particular distribution of magnitudes is found by multiplying the relative probabilities that each individual source is found at that specific magnitude,
\begin{equation}
\begin{aligned}
    &p(\textrm{distribution of } \mabs | \vec{\phi}, \envi) = \\
    &\prod_j^{n_i} \frac{f_i(M_j)\Phi _{\rm avg}(M_j,\vec{\phi})(1+\envi\epcvi(M_j,V_i))}{\int_{-\infty}^{M_{\rm lim}} f_i(M')\Phi _{\rm avg}(M',\vec{\phi})(1+\envi\epcvi(M',V_i))\textrm{d}M'},
\end{aligned}
\end{equation}
where $M_{\rm lim}$ is the magnitude limit of the survey.
Substituting in $n_{i,\rm exp} / V_i$ for the integral and plugging into equation~(\ref{eq:indvlikelihood}), we find
\begin{equation}
\begin{aligned}
    &p(D_i | \vec{\phi}, \envi) = \\
    &\frac{n_{i,\rm exp}^{n_i}}{n_i!} e^{-n_{i,\rm exp}} \times \\
    &\left(\frac{n_{i,\rm exp}}{V_i}\right)^{-n_i} \prod_j^{n_i} f_i(M_j)\Phi _{\rm avg}(M_j,\vec{\phi})(1+\envi\epcvi(M_j,V_i)).
\end{aligned}
\end{equation}
Inserting this into the full likelihood, taking the natural logarithm, and dropping terms that don't depend on $\vec{\phi}$ or $\vec{\delta}$ gives
\begin{equation}\label{eq:likelihood}
\begin{aligned}
    &\textrm{ln}\mathcal{L} \propto \\
    &\sum_i^{\snum} \left[ - n_{i,\rm exp} + \sum_{j}^{n_i} \left(\textrm{ln}\Phi_{\rm avg}(M_j,\vec{\phi}) + \textrm{ln}(1+\env_{i}\epcvi(M_j,V_i))\right)\right].
\end{aligned}
\end{equation}
We can then write the posterior as
\begin{equation}\label{eq:posterior}
    p(\vec{\phi},\vec{\delta} | \vec{D}) \propto \mathcal{L} \times p(\vec{\delta}) \times p(\vec{\phi}).
\end{equation}
Finally, we can marginalize over $\vec{\phi}$ or $\vec{\delta}$ to get $p(\vec{\delta} | \vec{D})$ or $p(\vec{\phi} | \vec{D})$, respectively.

A downside of considering the density of each field is the expanded dimensionality of the parameter space.
For example, given a luminosity function form that has $m$ parameters and a survey that has $\snum$ sub-fields, the posterior has a dimensionality of $m \times \snum$. However, each sub-field's $\env_i$ parameter is assumed to be independent, reducing the problem to $\snum$ different likelihoods each with dimensionality $m + 1$, a drastic reduction. These likelihoods can then quickly be combined to create the full posterior.

%%%%%%%%%%%%%%%%%%%%%%%%%%%%%%%%%%%%%%
\subsection{Combining fields}\label{sec:eff}

Unfortunately, sampling the posterior with many sub-fields can still be costly. For example, imagine a parallel survey with dozens of independent pointings, each of which provides a separate density and local luminosity function.
To alleviate this limitation, we next describe a method of combining multiple fields into a single `effective' field with $\env_{\rm eff}$ and $\epcveff$.

Combining fields is a trade-off. We reduce the dimensionality of the posterior, but we also lose the ability to measure individual field densities. This trade-off is well worth it when we only care about the density of a select number of fields in a survey, or in the following two cases: (i) the fields may be contiguous but at different depths (so that it might be useful to combine them in order to improve the environment measurement), or (ii) they may be widely separated and independent (but presumably shallow enough that no individual field will provide a robust environment anyway). We begin with the latter case.

%%%%%%%%%%%%%%%%%%%
\subsubsection{Independent fields}\label{sec:indp}

We begin with a method that is
especially useful in combining fields from a large parallel program, or from a handful of mosaics.

Take $\snum$ fields, each with a different effective volume curve $\veff$ (effective volume is the combined completeness/contamination function $f(\mabs)$ times total volume $\vol$). The fields are far apart from one another on the sky so are in independent environments.

Choosing a single field, the number of sources in a small magnitude bin is
\begin{equation}
    dN = V_{\rm eff}(\mabs)\Phi_{\rm avg}(\mabs)(1+\env\epcv(\mabs,\vol))d\mabs.
\end{equation}
At fixed $\mabs$, $dN$ is essentially a ‘measurable’ with average value $dN_{\rm avg} = V_{\rm eff}\Phi_{\rm avg}d\mabs $ and error in the measurement $\sigma_{\rm err} = V_{\rm eff} \Phi_{\rm avg} \epcv d\mabs $.
Combining the $\snum$ different fields, we would have $dN_{\rm tot} = \sum_i^{N_s} dN_i$ with the average value as $dN_{\rm tot, avg} = d\mabs V_{\rm eff, tot} \Phi_{\rm avg}$ and $V_{\rm eff, tot}(\mabs) = \sum_i^{N_s} V_{\rm eff, i}(\mabs)$. 
Via standard propagation of errors,
\begin{equation}
    \sigma_{\rm err, tot} = d\mabs V_{\rm eff, tot} \Phi_{\rm avg} \frac{\sqrt{(V_{\rm eff, 1}\epcvone)^2+(V_{\rm eff, 2}\epcvtwo)^2+...}}{V_{\rm eff, tot}}.
\end{equation}
We can then write the total expected number number of galaxies in this bin as
\begin{equation}
    dN_{\rm tot} = V_{\rm eff, tot}(\mabs)\Phi_{\rm avg}(\mabs)(1+\env_{\rm eff}\varepsilon_{cv,\rm eff}(\mabs)) d\mabs,
\end{equation}
with
\begin{equation}
    \varepsilon_{cv,\rm eff}(\mabs) = \frac{\sqrt{\sum_i^{N_s} [ V_{\rm eff, i}(\mabs)\epcvi(\mabs) ]^2}}{V_{\rm eff, tot}(\mabs)}
\end{equation}
as the effective bias function of the combined fields and $\env_{\rm eff}$ a normally-distributed random variable.
Not unexpectedly, the effective bias function is just the individual bias functions added in quadrature and weighted by volume.

Now, we may combine the data from $\snum$ independent fields and treat it as a single field using the ‘effective’ bias function and the total effective volume curve (which is just all of the individual curves added together).

Unfortunately, this method is technically correct only if all fields' effective volume curves $V_{\rm eff}$ have the exact same shape (though they may have different normalization). Fields that have very different effective volume curves should not be combined in this way. However, we find that fields that have consistent effective volume curves over the majority of their magnitude coverage can be combined this way with accurate results, even if one field goes $\sim$1 magnitude deeper than others.

%%%%%%%%%%%%%%%%%%%
\subsubsection{Contiguous fields}\label{sec:cont}

Now we turn to the case in which volumes are contiguous with each other (as may occur with ``wedding cake" surveys with embedded deep fields). Because the fields are contiguous, their densities are similar. 
Again, we take $\snum$ fields, each with a different effective volume curve $\veff$. For one field, the number of sources in a small magnitude range is
\begin{equation}
    dN = V_{\rm eff}(\mabs)\Phi_{\rm avg}(\mabs) \left[ 1+\frac{\delta_b}{\sigmapb}\epcv(\mabs) \right] d\mabs
\end{equation}
where this time, we have explicitly written the density $\env = \envreal/\sigmapb$ (see section~\ref{sec:lf}). Putting the fields together,
\begin{equation}
    dN_{\rm tot} = d\mabs \Phi_{\rm avg}  \left( V_{\rm eff, tot} + \sum_i^{N_s} V_{\rm eff, i}\frac{\delta_{b,i}}{\sigmapbi}\epcvi \right).
\end{equation}
We now make the assumption that $\delta_{b,i} = \delta_{\rm tot}$, i.e., the dark matter density is the same in all of the contiguous fields because they are near each other. This is only valid if $\sigmapb[{\rm within}] \ll 1$, where $\sigmapb^2[{\rm within}] = \sigmapb^2[{\rm smallest~field}] - \sigmapb^2[{\rm total~survey~volume}]$ is the 1-$\sigma$ fluctuation of dark matter density when zooming in from the entire contiguous survey to the smallest field.
Of course, in reality the different fields will not have the same density, though they will be correlated with each other. The assumption of a uniform density is a simplification useful for forecasting results; the inferred density will then be a weighted combination of the different parts of the field. If the true densities are of interest, our method can be extended to include these correlations
%Added a pointer to the later section
(see section~\ref{sec:measuringdensities}).

Re-writing again, we find
\begin{equation}
    dN_{\rm tot} = d\mabs V_{\rm eff, tot} \Phi_{\rm avg}  (1 + \frac{\delta_{\rm tot}}{\sigmapbtot}\sum_i^{N_s}\frac{V_{\rm eff, i}}{V_{\rm eff, tot}}\frac{\sigmapbtot}{\sigmapbi}\epcvi).
\end{equation}
In other words,
\begin{equation}
    dN_{\rm tot}(\mabs) = d\mabs V_{\rm eff, tot}(\mabs) \Phi_{\rm avg}(\mabs)  [ 1 + \env_{\rm eff}\varepsilon_{cv,\rm eff}(\mabs) ],
\end{equation}
with 
\begin{equation}
    \varepsilon_{cv,\rm eff}(\mabs) = \sum_i^{N_s}\frac{V_{\rm eff, i}(\mabs)}{V_{\rm eff, tot}(\mabs)}\frac{\sigmapbtot}{\sigmapbi}\epcvi(\mabs).
\end{equation}
and $\env_{\rm eff} = \delta_{\rm tot} / \sigmapbtot$ is just a normal random variable. This time, the effective bias function is weighted by both the effective volume and the individual fields' rms dark matter variation.

Again, we may combine the data from $\snum$ contiguous fields and treat them as a single field using the ‘effective’ bias function and the total effective volume curve. However, we must ensure our assumption of $\sigmapb[{\rm within}] \ll 1$ is valid. Through testing with our public cosmic variance calculator \pakidge~ \citep{Trapp2020}, $\sigmapb[{\rm within}]<0.2$ gives an effective bias function within ~10\% of the “correct” answer.

Finally, a mixture of independent and contiguous fields may be combined into an effective field by applying these methods one after another, always starting with the contiguous combinations.
The value of $\epcveff$ may also be used to forecast the aggregate effects of cosmic variance on any set of surveys before they are observed (given an estimate of their effective volumes).

To summarize, in this section we have introduced simple ways to combine several fields for a joint analysis. We emphasize that these  simplifications are largely for computational convenience; given the approximations already inherent to forecasting future surveys, they are certainly useful at this stage, but they may not be useful once the data are in hand.

%%%%%%%%%%%%%%%%%%%%%%%%%%%%%%%%%%%%%%
\subsection{Systematic offsets between fields}\label{sec:sysoff}

Unknown systematic offsets between fields complicates the measurement of their densities.
Our framework determines a field's density by comparison with other fields. If data from all fields are reduced in a consistent manner, any systematic errors in normalization would not affect this comparison. However, when applying our framework to systematically distinct datasets, a normalization offset could result in erroneous density measurements.

The parameter that holds these systematics is the effective volume $V_{\rm eff}$, which unfortunately cannot be measured, only modeled. Most often it is modeled by inserting artificial sources into simulations, taking a virtual observation of those simulations, and recording the number of correctly and incorrectly recovered sources \citep[e.g.][]{Finkelstein2015, Bouwens2015}. This method is robust and well developed, but the versions from different groups would ideally be tested on identical datasets to estimate their systematics and therefore uncertainties, which could then be accounted for in fitting.

Another source of potential systematic bias involves the galaxy selection process. In the same field, different selection criteria can identify different galaxy populations, which may have separate associated luminosity functions that should not be fit to with a single model.
We do not attempt to model such differences in our forecasts below, but they will be important to understand in future data. For simplicity, we assume that surveys are ``perfect" in that they can reliably identify all galaxies in the survey volume, aside from incompleteness at the faint end (modeled by $V_{\rm eff}$). In other words, we assume that the galaxy selection criteria used in the surveys do not ``miss" galaxy populations due to dust, old stars, etc. 
\citet{Kauffmann2020} find that this assumption is a reasonable one for predicting JWST survey results.

%%%%%%%%%%%%%%%%%%%%%%%%%%%%%%%%%%%%%%%%%%%%%%%%%%%%%%%%%%%%%%%%
%%%%%%%%%%%%%%%%%%%%%%%%%%%%%%%%%%%%%%%%%%%%%%%%%%%%%%%%%%%%%%%%

%%%%%%%%%%%%%%%%%%%%%%%%%%%%%%%%%%%%%%%%%%%%%%%%%%%%%%%%%%%%%%%%
%%%%%%%%%%%%%%%%%%%%%%%%%%%%%%%%%%%%%%%%%%%%%%%%%%%%%%%%%%%%%%%%

\section{Applications for JWST and beyond}\label{sec:applications}

In this section, we test our framework by simulating a set of upcoming JWST surveys and one Nancy Grace Roman Space Telescope survey in the range $6 \leq z \leq 12$. These simulations are designed to predict the performance of these surveys in regards to measuring the average luminosity function and field densities. It also serves as an example of how to apply our framework.

%%%%%%%%%%%%%%%%%%%%%%%%%%%%%%%%%%%%%%
\subsection{Simulating surveys}

We simulate upcoming surveys in a very simple manner, and as such, our forecasts cannot be seen as authoritative. However, they should provide reasonably accurate estimates of the expected precision of future measurements, and the general trends and qualitative relationships we find are robust and can be used as a tool to better plan future surveys. \citet{Kauffmann2020} contains a more detailed forecast for a smaller set of surveys.

We simulate 6 upcoming JWST surveys:
\begin{enumerate}
    \item the CEERS survey ($\sim$100 arcmin$^2$)
    \item the JADES survey ($\sim$236 arcmin$^2$)
    \item the PRIMER survey ($\sim$695 arcmin$^2$)
    \item the PANORAMIC parallel survey ($\sim$1500 arcmin$^2$)
    \item the WDEEP deep-field survey ($\sim$10 arcmin$^2$)
\end{enumerate}
We also simulate one Nancy Grace Roman Space Telescope survey: the Roman Supernova (RSN) survey ($\sim$32400 arcmin$^2$).
All survey features are described in Table \ref{tab:surveys}, though note that the RSN parameters are only estimates as the survey has not been finalized. We do not consider a comprehensive list of JWST galaxy surveys, choosing instead a representative sample. Other surveys will be useful (such as COSMOS-Webb) and more will be scheduled for later cycles.

\begin{table*}
	\centering
	\caption{Simulated survey parameters. The rms fluctuation of the dark matter density field at redshift 6 ($\sigmapb(z = 6)$) assumes $\Delta z = 1$; this value will differ at higher redshifts both due to the growth of structure and to the smaller volume at higher redshift for fixed angular area and $\Delta z$. The value of $\sigmapb(z = 6)$ and $\epcv(m = 26, z = 6)$ for these surveys illustrate that we are in the linear regime. Contrary to what one might infer from the growth of structure over cosmic time, $\epcv(m = 26, z = 6)$ \emph{increases} with redshift, as the volume decreases (with fixed area and $\Delta z$) and becomes less elongated (see Figure~\ref{fig:shapechange}).}
	\label{tab:surveys}
	\begin{tabular}{cccccc}
		\hline
		\hline
		Survey & Sub-Fields & App. mag lim. & Area & $\sigmapb(z = 6)$ & $\epcv(m=26,z=6)$\\
		 & & [mag] & [arcmin$^2$] & rms den. fluc. & galaxy bias\\
		\hline
        WDEEP & - & 30.75 & 10 & 0.055 & 0.32\\
        \hline
        CEERS & - & 28.97 & 100 & 0.038 & 0.25\\
        \hline
        JADES & & & 236 & - & -\\
              & North medium & 28.8 & 95 & 0.039 & 0.23\\
              & South medium & 28.8 & 95 & 0.039 & 0.23\\
              & South deep & 29.8 & 46 & 0.045 & 0.26\\
        \hline
        PRIMER & & & 695 & - & -\\
              & COSMOS shallow & 28.89 & 144 & 0.036 & 0.21\\
              & COSMOS medium & 29.11 & 108 & 0.038 & 0.23\\
              & COSMOS deep & 29.51 & 33 & 0.047 & 0.28\\
              & UDS shallow & 28.48 & 234 & 0.032 & 0.19\\ 
              & UDS medium & 28.89 & 175 & 0.034 & 0.21\\
        \hline
        PANORAMIC & 150x parallel & <29.5 & 1,500 & 0.055 each & 0.32 each\\
        \hline
        RSN & - & 28.8 & 32,400 & 0.005 & 0.03\\
        \hline
	\end{tabular}
\end{table*}

\begin{table}
	\centering
	\caption{Schechter parameters from \citet{Finkelstein2015}.}
	\label{tab:schparams}
	\begin{tabular}{cc}
		\hline
		\hline
		Parameter & Redshift Dependence \\
		\hline
        log$_{10}$[$\phi^*$] & $-1.58\pm0.3 - (0.31\pm0.07)z$ \\
        $\alpha$ & $-0.79\pm0.21 - (0.19\pm0.04) z$\\
        M$^*$ & $-20.27\pm0.42 - (0.12\pm0.09)  z$ \\
        \hline
	\end{tabular}
\end{table}

We simulate these surveys by first choosing a ``true'' luminosity function $\Phi_{\rm avg}$ to be a Schechter function with redshift-dependent parameters from \cite{Finkelstein2015} (see Table~\ref{tab:schparams}).
This choice does not strongly affect our results; we examine other ``true" luminosity functions in section~\ref{sec:limitations}.
For each field, we define a local luminosity function $\Phi_{\rm loc}$ (see eq.~\ref{eqn:philoc}) and draw a random value from a normal function for $\env$; $\epcv$ for each field is calculated with the python package
\pakidge~\citep{Trapp2020}, which uses the energy-regulated galaxy model of \citet{Furlanetto2017} and the halo mass function of \citet{Trac2015} converted into a conditional halo mass function with a scaling method from \citet{Tramonte2017}. 
Finally, we draw sources randomly from each field's ``observed'' luminosity function $\Phi_{\rm obs}$ (see eq.~\ref{eqn:phiobs}) with completeness curve $f(\mabs)$ by splitting $\Phi_{\rm obs}$ into small magnitude bins (0.05 magnitudes per bin) and drawing galaxies from a Poisson distribution around the expected value for each bin.

We create completeness curves $f(\mabs)$ by scaling the \HST~completeness curves from \cite{Finkelstein2015} to the appropriate magnitude limits. This procedure is only approximate but should serve as a reasonable stand-in for our purposes. All sources in the $z = 6$ bin are drawn from the $z = 6$ luminosity function, and likewise for the other redshifts. We use a redshift bin width of $\Delta z = 1$ for defining the volume of the surveys. In reality, the luminosity function is changing throughout that range, but that effect is small in practice (see Appendix~\ref{sec:deltaz}).

We consider many forthcoming surveys and combine them in various subsets (in the manner described in sections~\ref{sec:indp} and~\ref{sec:cont}). Many of the surveys themselves have subcomponents. For the purposes of computational efficiency in making our forecasts, we treat these surveys and their subcomponents in the following manner: 

\emph{(i)} The WDEEP, CEERS, and Roman SN fields are each treated as single independent volumes.

\emph{(ii)} For the JADES survey, we combine the  JADES South medium and deep fields contiguously. We combine this independently with the Jades North medium field.

\emph{(iii)} For the PRIMER survey, we combine the COSMOS shallow, medium, and deep fields contiguously, and we combine the UDS shallow and medium fields contiguously.

\emph{(iv)} PANORAMIC has approximately 150 separate pointings. We separate these into two groups by depth and combine each subset independently.

\begin{figure}
    \centering
    \includegraphics[width=0.5\textwidth]{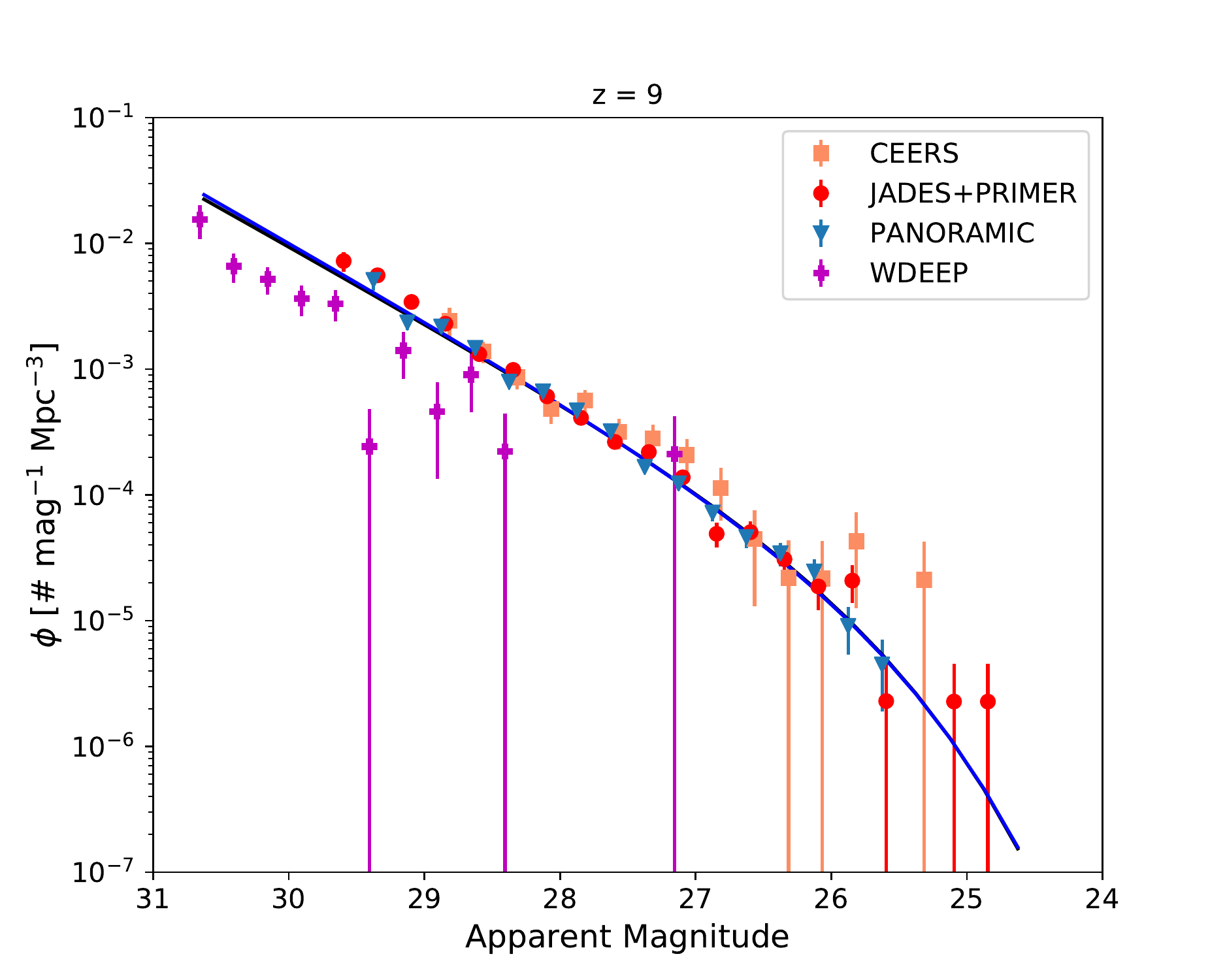}
    \caption{Data from one simulation of the CEERS+JADES+PRIMER+PANORAMIC+WDEEP surveys at $z = 9$. \textit{Black} line: input luminosity function; \textit{blue} line: best fit luminosity function. In this case, the WDEEP pointing happens to be a very under-dense region (see Figure~\ref{fig:post_Env}), but the ``true'' input luminosity function is still recovered.
    }
    \label{fig:data}
\end{figure}

When we combine many surveys, we make some additional combinations for computational convenience. In particular, we combine the PRIMER COSMOS, PRIMER UDS, JADES North, and JADES South fields together independently when all four fields are present in one survey combination. When the PANORAMIC survey is combined with other surveys, the two separate depth groups are combined independently. Figure~\ref{fig:data} shows the simulated data for one of the combinations of surveys.

We apply the framework described in section~\ref{sec:methods} to many simulated survey combinations to obtain Schechter parameter $\vec{\phi}$ posteriors for the average luminosity function and density $\vec{\delta}$ posteriors for the individual fields.
We are interested in the width of these posteriors as a measure of how well that survey combination constrains the luminosity function and field densities.
However, each instance of simulated surveys is affected by both cosmic variance and Poisson noise; different realizations of the same surveys will have differently-shaped posteriors. 
In order to get an understanding of how well a survey combination does on average, we repeat this whole process $N$ times for each survey. We choose $N$ such that the standard error in determining the average posterior width for each parameter is less than 10\%. 

In Table~\ref{tab:surveys}, the 2 right-most columns shows the rms linear density fluctuation $\sigmapb$ (at $z=6$) and galaxy bias (at apparent magnitude 26) for each of the fields. The values of $\sigmapb$ are all much less than unity, indicating we are within the linear regime of structure formation. $\epcv$ also remains well below unity, indicating our assumption of a gaussian local luminosity function (eq.~\ref{eqn:philoc}) is valid in these cases. However, as redshift increases, so does $\epcv$ (for fixed area and $\Delta z$); when $\epcv$ becomes larger than $\sim$0.5, the assumption of a gaussianity becomes worse. Fortunately, when $\epcv$ becomes larger than 0.5, Poisson noise typically takes over as the dominant source of uncertainty (see Figure~\ref{fig:shapechange}).

Finally, for a real-data comparison, we have applied our framework to the multi-field Hubble Space Telescope data set from \cite{Finkelstein2015} ($6\leq z\leq 9$ only) and the wide-field ground-based data from \cite{Bowler2015}. We describe the procedure fully in Trapp et al. (2021b, in prep.); here we include the results only for context in understanding the improvements JWST will provide.

%%%%%%%%%%%%%%%%%%%%%%%%%%%%%%%%%%%%%%
\subsection{Survey posteriors and validation}

In this section we verify that our framework recovers the input ``true'' luminosity function parameters and randomly drawn field densities. In the next sections, we explore how well the framework constrains these parameters.

\begin{figure*}
    \centering
    \includegraphics[width=5.0 in]{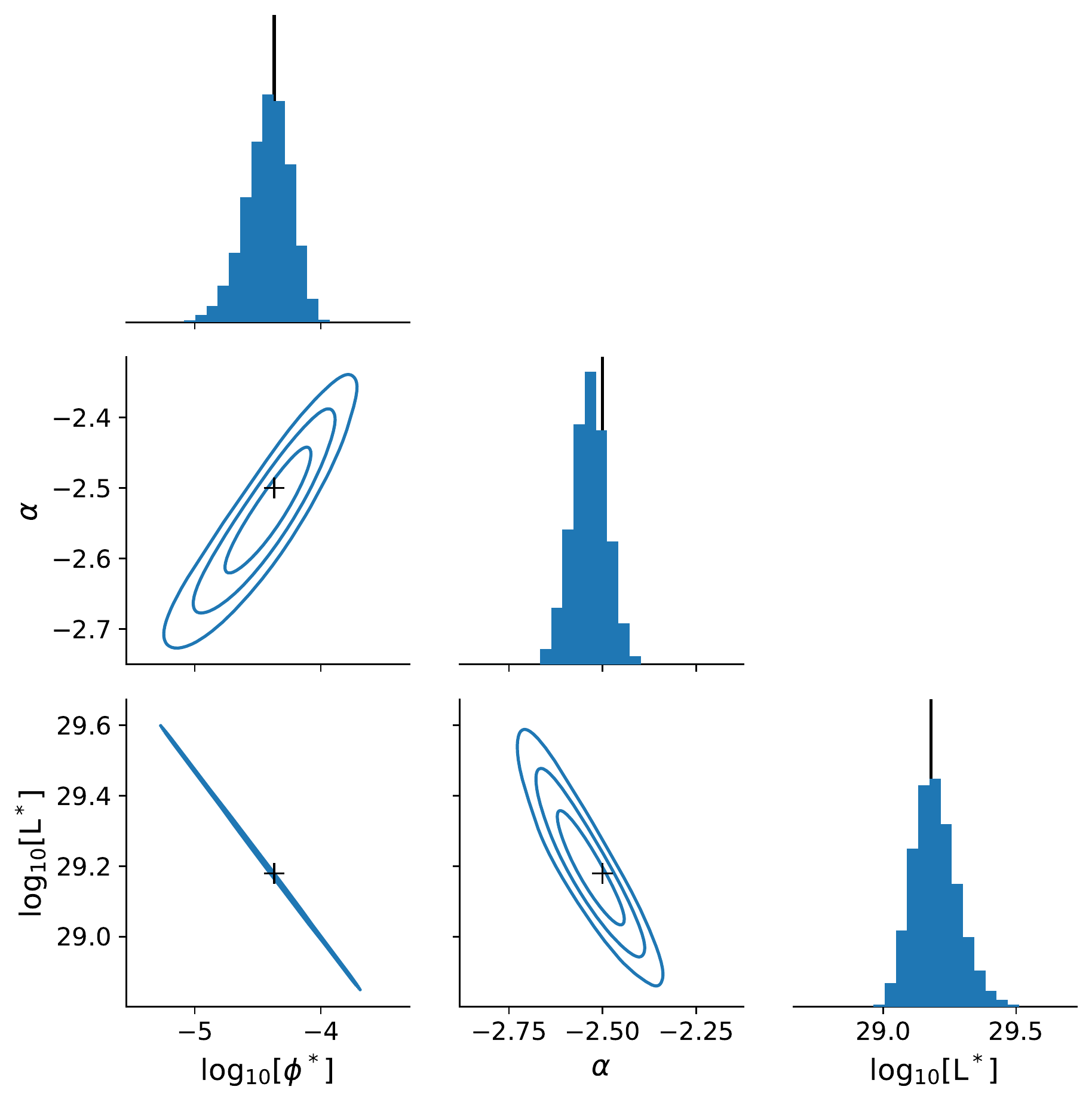}
    \caption{The Schechter parameter posterior (marginalized over all densities) for one simulation of the CEERS+JADES+PRIMER+PANORAMIC+WDEEP surveys at $z = 9$. The vertical black lines and the '+' marks show the ``true'' input values. At this redshift, the bright-end cutoff is barely covered by the data (see Figure~\ref{fig:data}), leaving a large (and tight) degeneracy between the normalization and characteristic luminosity.}
    \label{fig:post_Sch}
\end{figure*}

Figure~\ref{fig:post_Sch} shows the posterior of the Schechter parameters (marginalized over field densities) for the average luminosity function at $z = 9$ for one of the simulated survey combinations. 
The ``correct'' input values \citep[from][]{Finkelstein2015} are shown as vertical black lines. For this survey combination, there is a strong degeneracy between the normalization and the location of the faint-end cut-off (bottom left panel).

In general, our framework successfully recovers the correct input Schechter values except for cases where the surveys sample only the faint-end of the luminosity function, making the exponential cutoff impossible to localize. In these cases, the faint-end slope is still well recovered, but a two-parameter power-law model should be used instead of a full Schechter function. In the cases where the exponential cutoff is just barely covered by data, the fitting framework is biased towards choosing a larger normalization, shallower faint-end slope, and lower-luminosity exponential cutoff. This happens for the smaller-area ($A \lesssim 500$ arcmin$^2$) survey combinations at $z \gtrsim 9$.

In \citet{Trapp2020}, we found that fitting with cosmic variance in mind results in a less biased recovery of the luminosity function than the standard method used by e.g., \citet{Bouwens2015} and \citet{Finkelstein2015}. We do not explicitly compare our inference framework to others in this work, but we do point out that the luminosity function's shape changes more drastically with environment at higher redshift (see Figure~\ref{fig:shapechange}). Methods that assume the luminosity function has the same shape in all environments will lead to a worse fit.

\begin{figure*}
    \centering
    \includegraphics[width=5.0 in]{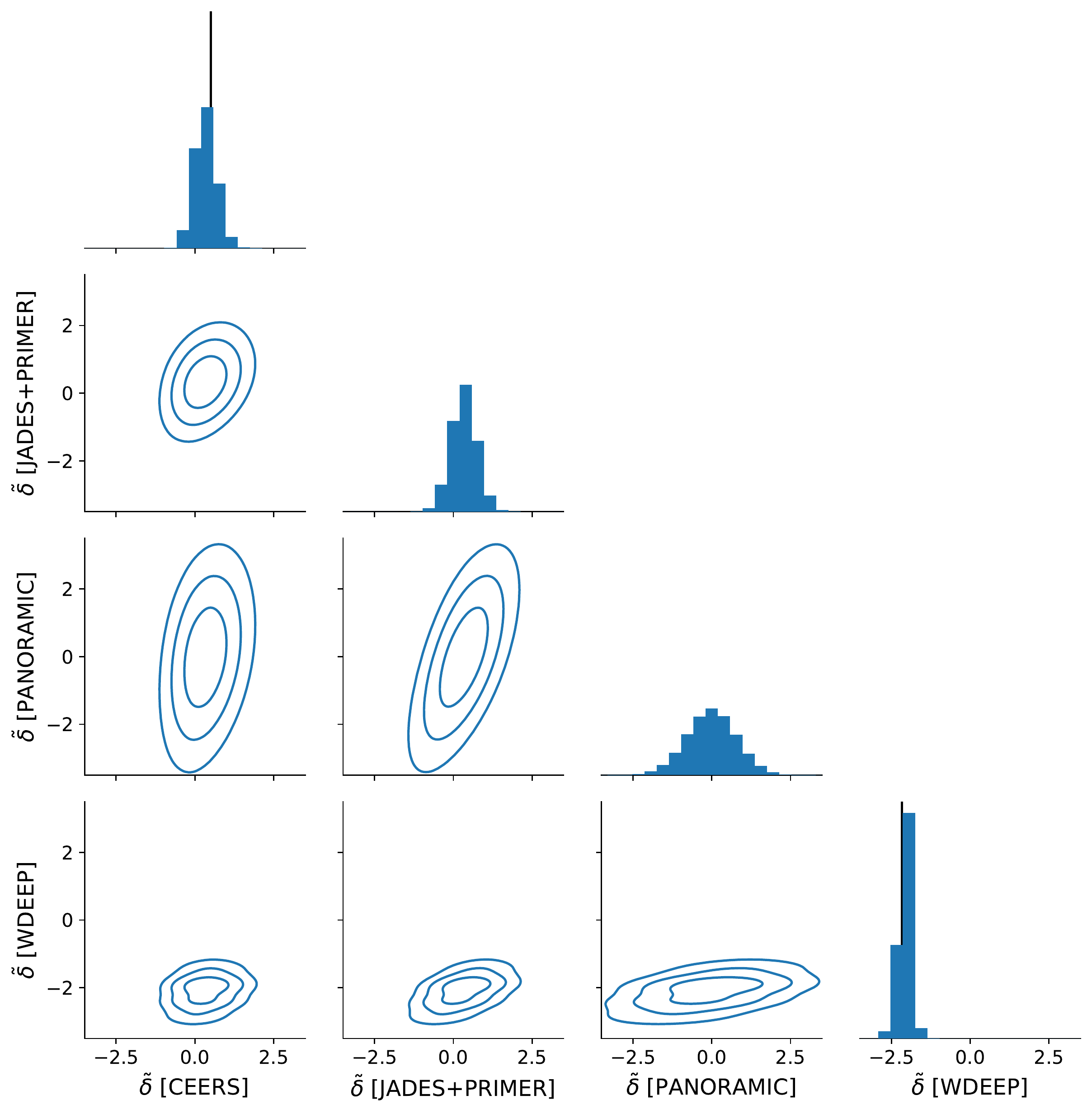}
    \caption{The density posterior (marginalized over all Schechter parameters) for one simulation of the CEERS+JADES+PRIMER+PANORAMIC+WDEEP surveys at $z = 9$. The vertical black lines on the histograms show the ``correct'' input density for CEERS and WDEEP. The ``correct'' densities for JADES+PRIMER and PANORAMIC are not shown as they are effective densities, not physical ones.}
    \label{fig:post_Env}
\end{figure*}

Figure~\ref{fig:post_Env} shows the posterior of the field densities at $z = 9$ for one of the simulated surveys.
The ``correct'' input values (drawn randomly for each field) are shown as vertical black lines for fields that have not been combined with others. Our framework's density posteriors consistently recover the correct input density values.

The resulting density posterior for PANORAMIC is not very informative. This is because PANORAMIC's density is an \emph{effective} density, a combination of multiple independent fields as described in section~\ref{sec:eff}. PANORAMIC's density is less well constrained than the other fields because it has a very small $\epcveff$ value, meaning a change in density is very difficult to distinguish from Poisson noise.

%%%%%%%%%%%%%%%%%%%%%%%%%%%%%%%%%%%%%%
\subsection{Measuring the luminosity function}\label{sec:measurelumfun}

A primary goal of JWST's extragalactic mission is to understand the physics and formation of high-$z$ galaxies. The luminosity function is the cornerstone of this effort and the key observable that models and simulations compare to. 

We now examine the constraining power of the simulated JWST surveys described in the previous sections.
Individual JWST surveys will be effective tools for measuring the luminosity function, but combining surveys gives even better constraints. We quantify those improvements by comparing the widths for the luminosity function parameters. We first compare the PRIMER, CEERS+JADES, and PANORAMIC individual performances. Then, we test various combinations of surveys using the PRIMER+CEERS+JADES survey as a baseline.

\begin{figure*}
    \centering
    \includegraphics[width=7.0 in]{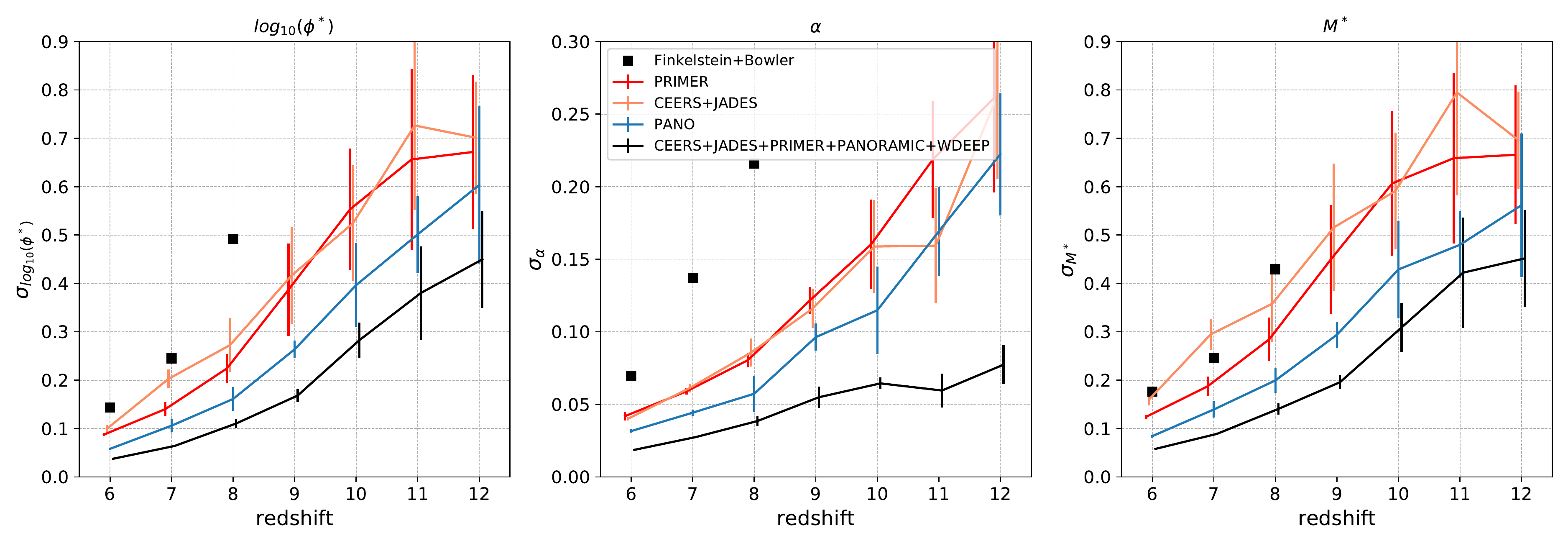}
    \caption{The 68.27\% confidence interval width of the Schechter parameters (averaged over many simulations) for different survey combinations as a function of redshift. The error bars represent the standard deviation of the posterior widths between simulations, which is due to Poisson noise as well as different random draws of the individual fields' densities. The black points are results from applying our framework to data from \citet{Finkelstein2015} and \citet{Bowler2015} for a comparison to HST's capabilities. Curves are shifted slightly in the $x$-direction to avoid overlapping error bars.}
    \label{fig:money_Sch_indv}
\end{figure*}

Figure~\ref{fig:money_Sch_indv} displays the 68.27\% confidence interval widths of the
posterior of the luminosity function parameters for redshifts $z = 6 - 12$ for the simulated PRIMER, CEERS+JADES, and PANORAMIC surveys. PRIMER and CEERS+JADES perform similarly, with PRIMER providing a slightly better measurement of the exponential cutoff due to its larger volume. Both PRIMER and CEERS+JADES would halve the uncertainty in the faint-end slope compared to HST at $z=8$ and significantly improve the normalization measurement. PANORAMIC by itself offers a significant improvement over PRIMER or CEERS+JADES (see Figure~\ref{fig:money_Sch_indv}). The combination of all JWST surveys is shown for comparison.

\begin{figure*}
    \centering
    \includegraphics[width=7.0 in]{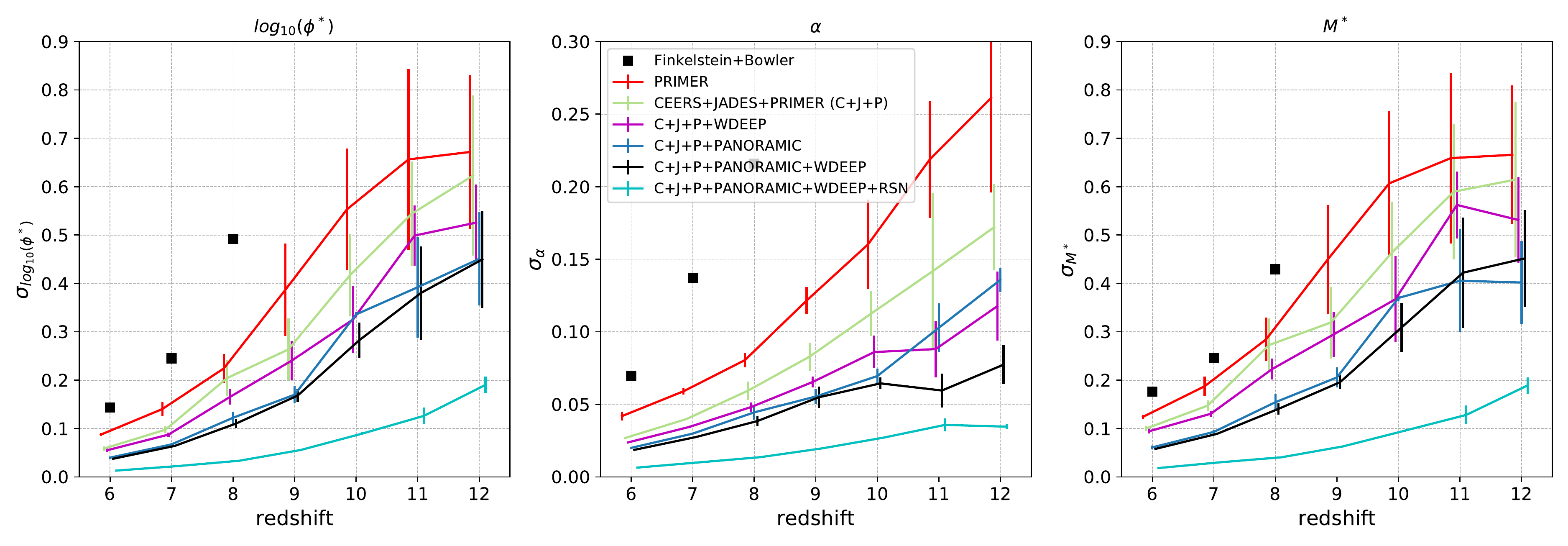}
    \caption{Same as Figure~\ref{fig:money_Sch_indv} but showing different combinations of surveys. The order of the lines in the legend (top to bottom) is the same as the order of the lines at $z = 8$ (top to bottom) for all three panels.}
    \label{fig:money_Sch_comb}
\end{figure*}

Figure~\ref{fig:money_Sch_comb} shows several additional survey combinations. 
Combining CEERS+JADES with PRIMER (C+J+P) offers a significant improvement (green line). Adding WDEEP to C+J+P improves the measurement of the faint-end slope much more than the other two parameters, because it goes significantly deeper than any other survey. Adding just PANORAMIC to C+J+P improves the constraints on all parameters. 
When combining all JWST surveys, we (unsurprisingly) get the best constraints (black line).

With all surveys combined, JWST will measure the faint-end slope at $z = 12$ to similar precision as HST measures at $z = 6$--$7$. JWST will measure the normalization and exponential cutoff at $z = 12$ to similar precision as HST measures at $z = 8$.

We add one additional further-future survey, the Roman SN survey, which will vastly improve the measurement of the exponential cutoff and normalization due to its large volume. Perhaps surprisingly, the faint-end slope is also much better measured. With the degeneracy between the exponential cutoff and normalization broken, the faint-slope is easier to pin down with the JWST surveys. This effect highlights the importance of wide-field searches for the brightest objects \citep[see also][]{Kauffmann2020}.

One may wonder how much the parallel nature of PANORAMIC helps in the constraints. To test this, we also considered an artificial survey that is identical to the PANORAMIC survey but in two contiguous fields (one incorporating the deeper pointings and one the shallower ones). The constraints are nearly identical to those from the full PANORAMIC, likely due to the fact that the area is so large (1500 arcmin$^2$) that cosmic variance's effects are small even if it were a contiguous field. However, the parallel nature of PANORAMIC provides other unique opportunities (see section~\ref{sec:galphy}).

The posterior widths presented here are marginalized over field densities. Fitting a luminosity function to data without fully incorporating cosmic variance leads to larger posterior widths and, in some cases, biased results \citep[see][]{Trapp2020}.

%%%%%%%%%%%%%%%%%%%%%%%%%%%%%%%%%%%%%%
\subsection{Measuring field densities}\label{sec:measuringdensities}

We have now seen that, in the JWST era, galaxy abundances throughout the Cosmic Dawn will be measured to unprecedented precision. But our new method also opens an entirely new window onto that era, by providing measurements of the large-scale environments of survey fields. At least in principle, such environmental measurements will allow us to associate galaxies with their ionization environments, their future descendant populations, and to conduct searches for how galaxy evolution depends on those environments.

\begin{figure*}
    \centering
    \includegraphics[width=6.0 in]{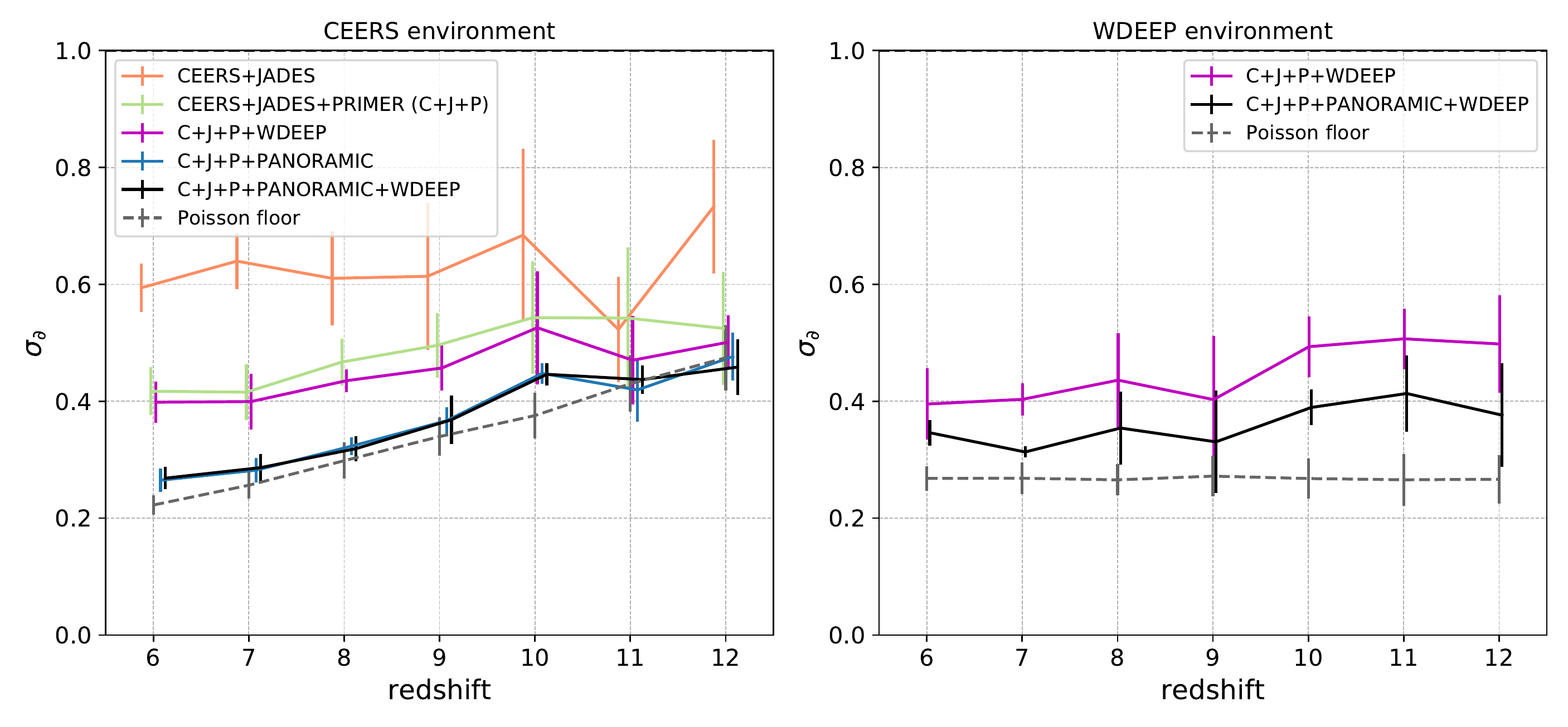}
    \caption{The 68.27\% confidence interval widths of the CEERS and WDEEP field densities (averaged over many simulations) for different survey combinations as a function of redshift. The error bars represent the standard deviation of the width in the posterior between runs, which is due to Poisson noise as well as different random draws of the individual fields' densities. The dashed black line shows the maximum accuracy in determining the field density (see section~\ref{sec:measuringdensities}). Curves are shifted slightly in the $x$-direction to avoid overlapping errorbars.
    %AT: Not sure what is up with CEERS+JADES at z = 11; I'll check it out.
    %AT: It appears to just be a fluctuation.
    }
    \label{fig:money_Env}
\end{figure*}

The density of a single field can only be determined in comparison to other fields. As more fields are added, the measurement of the average luminosity function's normalization gets better, and so too does the measurement of an individual field's density. This effect can be seen in the density posteriors for our fields. Figure~\ref{fig:money_Env} shows the 68.27\% confidence interval width of the marginalized posterior for the CEERS and WDEEP densities as a function of redshift for various combinations of surveys. The errorbars reflect the natural variance in the posterior widths from the effects of sample variance and Poisson noise. 

With CEERS and JADES alone, the CEERS environment can be measured with a posterior width\footnote{$\env$ is in units of standard deviations away from mean density. An over-density of $\env$ means there are a factor of $(1+\epcv \env)$ more/fewer galaxies than average. The prior on $\env$ is a normal with width $\sigma_{\env} = 1$.} of $0.6$--$0.7$. Combining CEERS+JADES with PRIMER improves this measurement significantly. Adding a large parallel survey like PANORAMIC does an even better job at improving the density measurement of the CEERS field. Adding a deep field like WDEEP to C+J+P does not greatly increase the measurement of the environment as WDEEP is a small volume with a large bias factor and thus does not help constrain the normalization as well as other surveys. When combining all surveys' data, JWST can measure the density of a 100 arcmin$^2$ field (like CEERS) to a precision of $0.3$ and 0.6 at $z = 6$ and $12$, respectively, and a deep 10 arcmin$^2$ field (like WDEEP) to a precision of $0.35$ or $0.45$ at $z = 6$ and $12$, respectively.

Clearly, more data result in tighter constraints on the density of any given field. However, the results appear to asymptote as more and more surveys are added, indicating there is a fundamental limit to measuring a field's density. That fundamental limit is Poisson noise, plotted as the gray dashed line.
This ``Poisson floor'' is calculated by simulating a survey field with Poisson noise and some density $\env$ drawn from a normal distribution. Then, the posterior of $\env$ is calculated in a similar manner to the posterior calculation in section~\ref{sec:bayes}, but assuming perfect knowledge of the luminosity function. The width of that posterior represents the best-case determination of the density, which we refer to as the `Poisson floor'. This floor depends slightly on the field's actual density (which was drawn randomly), and the variance from that effect is represented by the error-bars in the Poisson floor. The Poisson floor also depends weakly on the assumed luminosity function and strongly on the completeness function of the survey.

\begin{figure}
    \centering
    \includegraphics[width=0.5\textwidth]{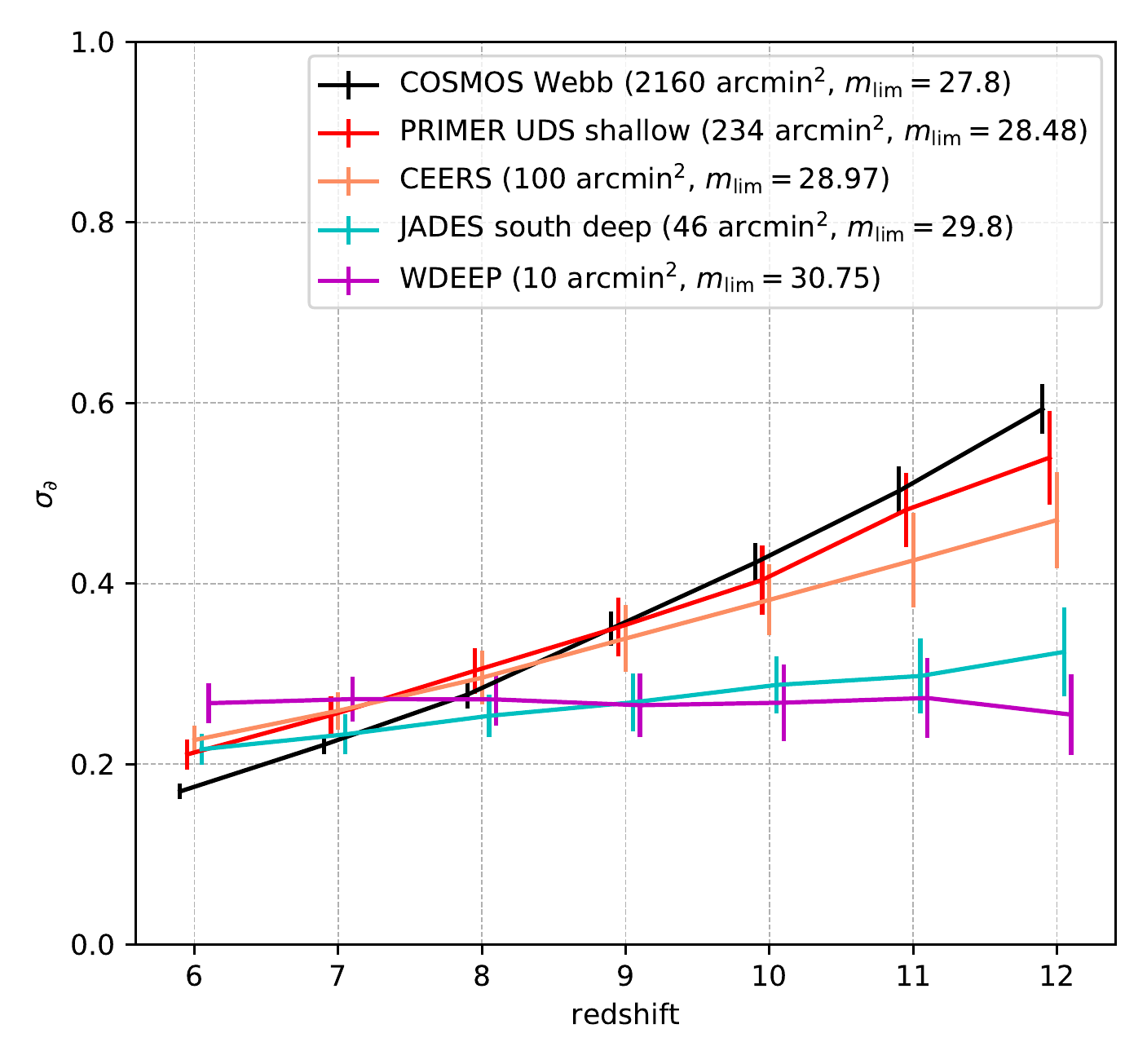}
    \caption{The best possible precision in measuring the density for various contiguous JWST fields (see section~\ref{sec:measuringdensities}). This limit is due to Poisson noise only; the luminosity function is assumed to be perfectly known. The prior on the density is $\sigma_{\partial} = 1$. A deeper survey has a lower and flatter curve (at fixed area), and a larger area survey has a lower curve (at fixed depth). Curves are shifted slightly in the $x$-direction to avoid overlapping errorbars.}
    \label{fig:poissfloor}
\end{figure}

We plot the Poisson floor for a variety of our fields in Figure~\ref{fig:poissfloor}. From testing with our framework, we find JWST will be able to measure all of its fields' densities to near this limit when multiple surveys are combined.

%%%%%%%%%%%%%%%%%%%%%%%%%%%%%%%%%%%%%%
\subsection{Testing galaxy physics in different environments}\label{sec:galphy}

A large parallel survey like PANORAMIC will probe a wide range of densities (150 independent pointings). After measuring the global luminosity function as in section~\ref{sec:measurelumfun}, we can go back and measure the density of each of the PANORAMIC fields using equations~(\ref{eq:likelihood}) and~(\ref{eq:posterior}). This time, the luminosity function prior is determined by the earlier global fit, over whose errors we can marginalize.\footnote{An individual PANORAMIC field contributes so little to the global fit of the luminosity function that it can safely be considered independent, as this description is assuming.}

While an individual PANORAMIC pointing will have an uncertainty in its density ($\sigma_{\env} \sim$0.4-0.8, from testing), they can be sorted from least to most dense. One can then search for correlations between dark matter density and galaxy properties, ionization history, etc.

A similar procedure could be applied to large contiguous fields like PRIMER, with the added complication that the field densities are correlated, being physically next to one-another.

We do caution the reader that we have not incorporated uncertainties in the galaxy models themselves into our inference framework; in particular, we assume a perfectly known galaxy bias function as expressed by $\epcv$. In the future, uncertainties in the galaxy model should be marginalized over. Fortunately, they are not large (see section \ref{sec:lf}), and at least in principle can be separated because the bias functions depend on halo mass (and hence galaxy luminosity).

%%%%%%%%%%%%%%%%%%%%%%%%%%%%%%%%%%%%%%
\subsection{Limitations and future improvements}\label{sec:limitations}

In our simulations, we use a luminosity function calibrated at low redshift ($z \approx 5-9$) and extrapolated to higher redshift \citep[][see our Table~\ref{tab:schparams}]{Finkelstein2015}. We cannot know if this extrapolation holds beyond the redshift where we have data, and we will not know until JWST and other observatories make their measurements. However, this extrapolation has reported uncertainties, so we re-do our analysis for a more slowly-evolving and more rapidly-evolving case based on those uncertainties. The slowly-evolving case has a shallower (increased) slope with redshift and decreased $y$-intercept for all Schechter parameters, and the rapidly-evolving case has steeper (decreased) slope and increased $y$-intercept (see Table~\ref{tab:schparams}).

The slowly-evolving case is similar to the fiducial case when measuring log$_{\rm 10}(\phi^*)$ and $M^*$, but it generally has a worse determination of the faint-end slope $\alpha$ with up to $\sim$30\% wider posteriors, depending on the combination of surveys and the redshift considered.
The rapidly-evolving case is similar to the fiducial case when measuring $\alpha$, but it generally has a worse determination of log$_{\rm 10}(\phi^*)$ and $M^*$ with up to $\sim$40\% wider posteriors, again depending on the survey combination and redshift.
These variations are usually smaller in scale than the variations between individual trials of the surveys, meaning our predictions are relatively robust to different luminosity functions.

As discussed in section~\ref{sec:lf}, the galaxy bias function $\epcv$ has an estimated uncertainty of $\sim$25\%. To test the effects of a different ``true'' bias function, we re-do our analysis again with a 25\% lower/higher value for $\epcv$ (when generating simulations). We find that we recover log$_{\rm 10}(\phi^*)$ and $M^*$ to similar precision as the fiducial case, but generally have a worse determination of the faint-end slope $\alpha$ with up to $\sim$25\% wider posteriors depending on survey combination and redshift.

One of the benefits of our framework is its potential for constraining the bias function itself from the data by treating it as a free parameter with a prior based on its estimated uncertainty and marginalizing the posterior space over all other parameters. However, we do not explore that potential constraining power in this work.

In our calculation of the posterior, we assume that all measurements of galaxy magnitudes (real and simulated) have no uncertainty. Taking these uncertainties into account would make the measurement of the bright end of the luminosity function (where it is steepest) more realistic, but also more uncertain and require more computation time. We plan to explore this trade-off in the future.

We have described the conditional luminosity function (eq.~\ref{eqn:philoc}) as being gaussian with respect to the density $\env$. However, as $\epcv$ approaches unity, the conditional luminosity function (at fixed magnitude) is better described by a log-normal or gamma function with standard deviation equal to $\epcv$ \citep[see Figure 6 in][]{Trapp2020}. To correct for this, we switch from gaussian to log-normal when $(1+\env\epcv) \leq 0$ (when the number density becomes negative). This correction is approximate; a full treatment requires the full conditional luminosity function \citep[see][]{Trapp2020}. However, this problem will affect only the brightest sources in the smallest volumes where Poisson noise is typically dominating anyway.

We use a very simple estimate of the completeness function for the JWST surveys. Actual completeness functions will require detailed knowledge of the telescope's performance and sophisticated simulations \citep[see][for detailed predictions of JWST's completeness]{Kauffmann2020}. If our estimates are off in normalization, that will increase the effect of Poisson noise but should not change the qualitative results much. If the shape of our completeness function is wrong, it will likely only affect the faint end of the luminosity function where the completeness function deviates significantly from its maximum value. 
The completeness can also moderately affect the Poisson floor which is sensitive to the faint-end of the luminosity function where the largest number of sources are found.

%%%%%%%%%%%%%%%%%%%%%%%%%%%%%%%%%%%%%%%%%%%%%%%%%%%%%%%%%%%%%%%%
%%%%%%%%%%%%%%%%%%%%%%%%%%%%%%%%%%%%%%%%%%%%%%%%%%%%%%%%%%%%%%%%

%%%%%%%%%%%%%%%%%%%%%%%%%%%%%%%%%%%%%%%%%%%%%%%%%%%%%%%%%%%%%%%%
%%%%%%%%%%%%%%%%%%%%%%%%%%%%%%%%%%%%%%%%%%%%%%%%%%%%%%%%%%%%%%%%
\section{Conclusions}\label{sec:conclusion}

We develop a framework that calculates the posteriors of the average galaxy luminosity function and field densities simultaneously. 
We also develop a method to combine independent and/or contiguous fields into ``effective'' fields with their own bias functions, reducing computation time and providing an estimate for the effects of cosmic variance on a complicated set of surveys. 

We find that JWST will improve and extend measurements of the luminosity function, with precision at $z = 12$ roughly equal to what HST is capable of at $z = 8$. With all early-cycle JWST programs combined, we expect to measure the normalization of the luminosity function to a precision of $0.05$ and $0.6$ dex at $z = 6$ and $12$, the faint-end power-law index to a precision of $0.03$--$0.10$ over the same redshift interval, and the characteristic magnitude to a precision of $0.08$ and $0.55$ at $z = 6$ and $12$ (see Figure~\ref{fig:money_Sch_comb}).

Large-area surveys are most important for breaking the degeneracy between the normalization and the bright-end cutoff. For the highest redshifts ($z \gtrsim 9$), a single power-law fit will likely be better than a Schechter fit even if the underlying population is a Schechter function, as probing the bright-end cutoff will require Roman-sized survey areas.

Parallel surveys are subject to less cosmic variance than contiguous surveys, but at sufficiently large area, contiguous surveys are mostly unaffected by cosmic variance.

When combined, early-cycle JWST galaxy surveys will be able to measure the normalized densities $\env$ of individual survey fields with a 68.27\% confidence interval of $\sim$0.4 for deep fields such as WDEEP, and $\sim$0.5 for larger fields like CEERS, nearly their theoretical maximum precision (see Figures~\ref{fig:money_Env} and~\ref{fig:poissfloor}). 

For these purposes, consistent data reduction between different surveys will be crucial, as the determination of field density depends on the relative normalization between surveys.

Measurements of large-scale environments will open new opportunities to study galaxy evolution during the Cosmic Dawn. Most importantly, reionization is driven by the underlying matter distribution (through the biased formation of galaxies; \citealt{Furlanetto2004}), so density measurements can also provide estimates of the local ionization state (albeit in a model-dependent fashion). Interestingly, the transverse scales of many anticipated fields are comparable to the ionized bubbles that appear through reionization -- for example, a comparison to \citet{Lin2016} and \citet{Davies2021} shows that WDEEP and PANORAMIC fields are comparable in extent to bubbles halfway through reionization, while CEERS and JADES fields are close to the sizes in the later stages.  Studies across such fields will allow exploration of the effects of reionization on the local galaxy population -- which has long been expected (e.g., \citealt{Thoul1996, Iliev2007, Noh2014}) but never observed directly. Density measurements will also allow studies of more conventional effects of environment on galaxy evolution and the quantitative association of high-$z$ environments with their descendants. 

Finally, density measurements allow targeted exploration of unusual environments (such as protoclusters) whose histories have long been studied. While our method provides only a first step toward such ambitious goals, it demonstrates that such inferences will soon be possible with forthcoming observational programs.

Our forecasts demonstrate the transformative potential of future galaxy surveys in understanding the Cosmic Dawn on both large and small scales. While harvesting this information will require improvements to the model (such as localizing galaxies in the radial direction, accounting for correlations between neighboring volumes, and incorporating uncertainties in the galaxy physics itself), the enormous potential provides strong motivation for such future efforts.

%%%%%%%%%%%%%%%%%%%%%%%%%%%%%%%%%%%%%%

%%%%%%%%%%%%%%%%%%%%%%%%%%%%%%%%%%%%%%%%%%%%%%%%%%%%%%%%%%%%%%%%
%%%%%%%%%%%%%%%%%%%%%%%%%%%%%%%%%%%%%%%%%%%%%%%%%%%%%%%%%%%%%%%%

%%%%%%%%%%%%%%%%%%%%%%%%%%%%%%%%%%%%%%%%%%%%%%%%%%%%%%%%%%%%%%%%
%%%%%%%%%%%%%%%%%%%%%%%%%%%%%%%%%%%%%%%%%%%%%%%%%%%%%%%%%%%%%%%%
\section*{Acknowledgements}

We thank C.~Williams for sharing details on the PANORAMIC survey, and S.L.~Finkelstein and R.A.A.~Bowler for making available their galaxy survey data. We would also like to thank Richard H. Mebane, F.~Davies, and J.~Mirocha for helpful conversations, and Jon K. Zink for providing expertise on statistical methods. 

This work was supported by the National Science Foundation through award AST-1812458. In addition, this work was directly supported by the NASA Solar System Exploration Research Virtual Institute cooperative agreement number 80ARC017M0006. We also acknowledge a NASA contract supporting the ``WFIRST Extragalactic Potential Observations (EXPO) Science Investigation Team" (15-WFIRST15-0004), administered by GSFC. J.~Y. thanks the UCLA Department of Physics \& Astronomy for support during its 2020 Undergraduate Summer Research Program.

\textit{Software used:} This work makes use of iPython \citep{Perez2007} and the following Python packages: NumPy \citep{Walt2011}, SciPy \citep{Virtanen2020}, Matplotlib \citep{Hunter2007}, and pandas \citep{McKinney2010}.

%%%%%%%%%%%%%%%%%%%%%%%%%%%%%%%%%%%%%%%%%%%%%%%%%%%%%%%%%%%%%%%%
%%%%%%%%%%%%%%%%%%%%%%%%%%%%%%%%%%%%%%%%%%%%%%%%%%%%%%%%%%%%%%%%

%%%%%%%%%%%%%%%%%%%%%%%%%%%%%%%%%%%%%%%%%%%%%%%%%%%%%%%%%%%%%%%%
%%%%%%%%%%%%%%%%%%%%%%%%%%%%%%%%%%%%%%%%%%%%%%%%%%%%%%%%%%%%%%%%
\section*{Data Availability}
No new data were generated or analysed in support of this research.

%%%%%%%%%%%%%%%%%%%%%%%%%%%%%%%%%%%%%%%%%%%%%%%%%%%%%%%%%%%%%%%%
%%%%%%%%%%%%%%%%%%%%%%%%%%%%%%%%%%%%%%%%%%%%%%%%%%%%%%%%%%%%%%%%



\bibliographystyle{mnras}
\bibliography{me} % if your bibtex file is called example.bib

%\begin{thebibliography}{99}
%\end{thebibliography}

%%%%%%%%%%%%%%%%%%%%%%%%%%%%%%%%%%%%%%%%%%%%%%%%%%

%%%%%%%%%%%%%%%%% APPENDICES %%%%%%%%%%%%%%%%%%%%%

\appendix

%%%%%%%%%%%%%%%%%%%%%%%%%%%%%%%%%%%%%%

\section{Choosing Optimal Redshift Bins} \label{sec:deltaz}

\begin{figure}
    \centering
    \includegraphics[width=0.5\textwidth]{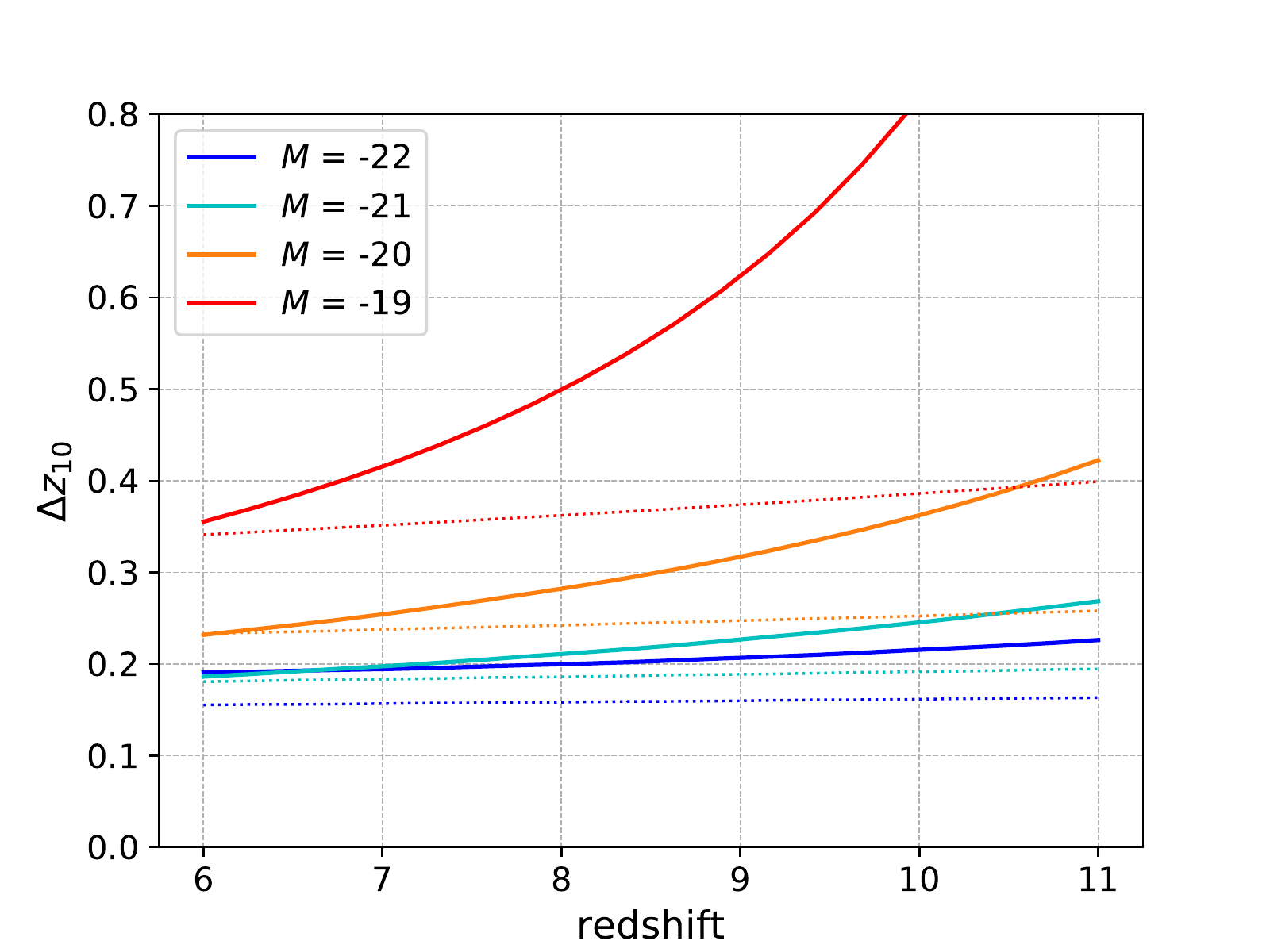}
    \caption{The change in redshift (towards lower redshift) that corresponds to a 10\% growth in the luminosity function. The dotted lines are the same but for the more slowly-evolving luminosity function (see section~\ref{sec:limitations}).}
    \label{fig:deltaz10}
\end{figure}

It is ideal to measure the luminosity function in as narrow a redshift range as possible, as the luminosity function is constantly evolving. However, it is often necessary to use large redshift bins to get a sufficient number of galaxies for fitting. How large then is too large?
We define a function $\Delta z_{10}(z,\mabs)$: the redshift change that corresponds to a $10$\% growth in the luminosity function at magnitude $\mabs$. Figure~\ref{fig:deltaz10} shows $\Delta z_{10}(z,\mabs)$ for a selection of magnitudes.
Figure~\ref{fig:deltaz10} would suggest that $\Delta z$ should be below one, especially for the bright end of the luminosity function. However, this can be mitigated by averaging over a wider bin.

We calculate the volume averaged luminosity function over the redshift range $\Delta z$ centered at $z_{\rm c}$ and then find the corresponding redshift with the same luminosity function value, defining that redshift as the `effective' redshift of the range $z_{\rm eff}(\mabs, \Delta z, z_{\rm c})$. The `effective' redshift is a function of magnitude, and it is what \emph{should} be used as the true center redshift, not $z_{\rm c}$. For survey estimates, the key question is the error introduced by using the central redshift instead of this (magnitude-dependent) effective value $z_{\rm eff}$.

\begin{figure}
    \centering
    \includegraphics[width=0.5\textwidth]{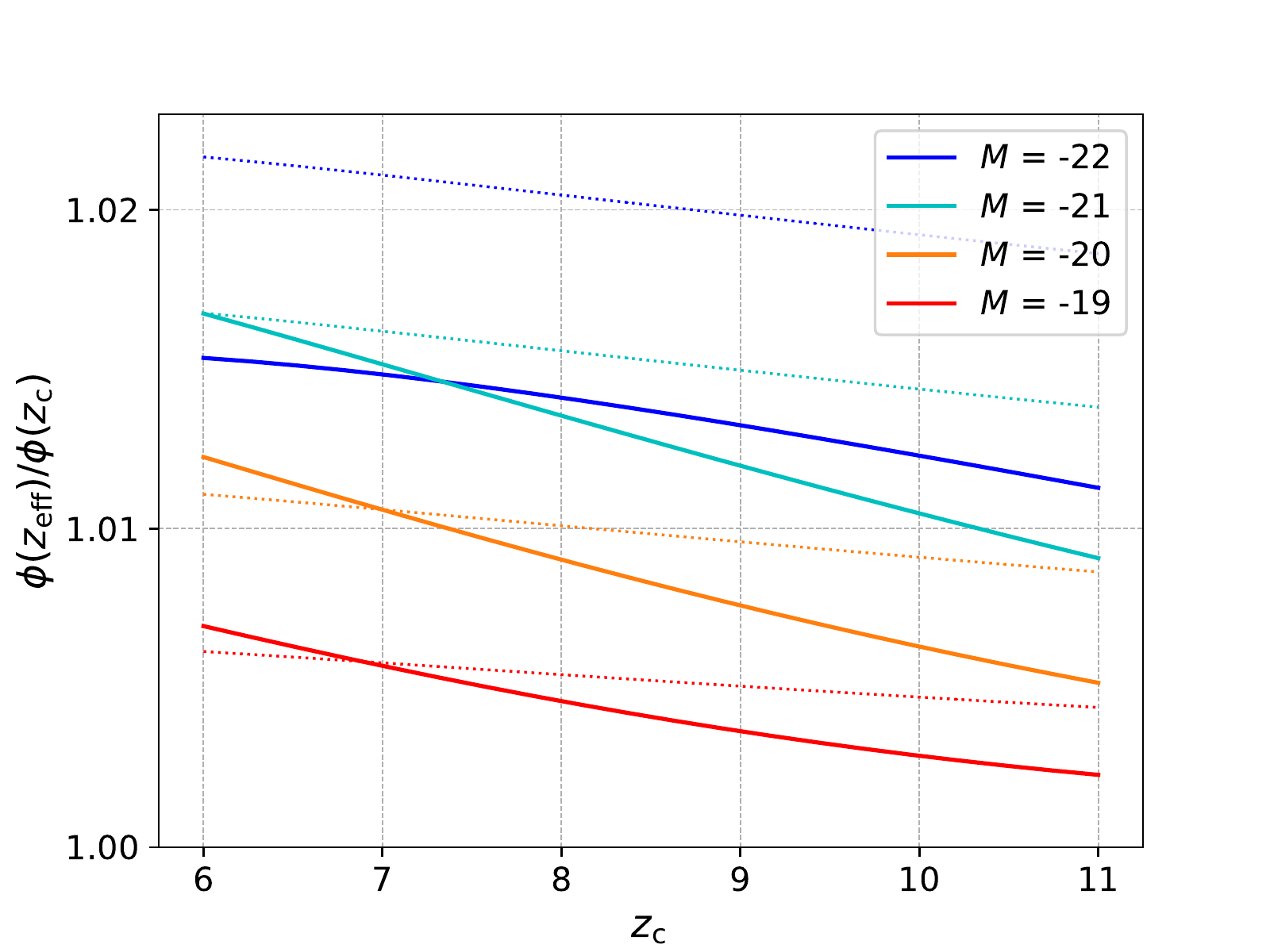}
    \caption{The ratio of the luminosity function at $z_{\rm eff}$ (the effective center of a bin centered at $z_{\rm c}$ with $\Delta z = 1$) and the luminosity function at $z_{\rm c}$. The dotted lines are the same but for the more slowly-evolving luminosity function (see section~\ref{sec:limitations}).}
    \label{fig:deltaz}
\end{figure}

We plot the ratio between the luminosity function at $z_{\rm eff}$ and $z_{\rm c}$ for a variety of magnitudes and redshifts in Figure~\ref{fig:deltaz}. We find no significant difference ($\lesssim$2\%) when using a redshift bin width of $\Delta z = 1$. Increasing $\Delta z$ to 2 results in deviations increasing to $\sim$7\% for the brightest sources.

These calculations depend on the shape and evolution of the luminosity function. By default, we use the Schechter function fit from \citet{Finkelstein2015}. The dotted lines in Figures~\ref{fig:deltaz10} and~\ref{fig:deltaz} are the same but for the more slowly-evolving luminosity function (see section~\ref{sec:limitations}); this effect does not strongly depend on luminosity function choice.

Because the errors introduced by the redshift binning are much smaller than our expected uncertainties, we use a constant $z_c$ in the main text.

%%%%%%%%%%%%%%%%%%%%%%%%%%%%%%%%%%%%%%%%%%%%%%%%%%

% Don't change these lines
\bsp	% typesetting comment
\label{lastpage}
\end{document}